\documentclass[prb,aps,twocolumn,showpacs,superscriptaddress,floatfix]{revtex4}
\usepackage{graphicx}
\usepackage{wasysym}

\newcommand{\beq}{\begin{equation}}
\newcommand{\eeq}{\end{equation}}
\newcommand{\bea}{\begin{eqnarray}}
\newcommand{\eea}{\end{eqnarray}}
\newcommand{\cI}{{\cal I}}
\newcommand{\cS}{{\cal SF}}

\newcommand{\be}{\begin{equation}}
\newcommand{\br}{{\bf r}}

\newcommand{\ee}{\end{equation}}
\newcommand{\bk}{{{\bf{k}}}}

\newcommand{\ra}{\rangle}
\newcommand{\la}{\langle}

\begin{document}

\title{Exotic phase diagram of a cluster charging model of
bosons on the kagome lattice}
\author{Sergei V. Isakov}
\affiliation{Department of Physics, University of Toronto, Toronto,
Ontario M5S 1A7, Canada}
\author{Arun Paramekanti}
\affiliation{Department of Physics, University of Toronto, Toronto,
Ontario M5S 1A7, Canada}
\author{Yong Baek Kim}
\affiliation{Department of Physics, University of Toronto, Toronto,
Ontario M5S 1A7, Canada}
\affiliation{School of Physics, Korea Institute for Advanced Study,
Seoul 130-722, Korea}

\date{\today}
\begin{abstract}
We study a model of hard-core bosons on the kagome lattice with 
short-range hopping ($t$) and repulsive interactions ($V$). This model
directly maps on to an easy-axis $S=1/2$ XXZ model on the kagome
lattice and is also related, at large $V/t$,
to a quantum dimer model on the triangular lattice.
Using quantum Monte Carlo (QMC) numerics, we map out the phase diagram
of this model at half-filling. At $T=0$, we 
show that this model exhibits a superfluid phase at small $V/t$ and an
insulating phase at large $V/t$, separated 
by a continuous quantum phase transition at $V_c/t \approx 19.8$. 
The insulating phase at $T=0$
appears to have no conventional broken symmetries, and is thus a uniform
Mott insulator (a `spin liquid' in magnetic language).
We characterize this insulating phase as a uniform $Z_2$ 
fractionalized insulator from the topological order in the ground state 
and estimate its vison gap.  Consistent with this 
identification, there is no apparent thermal phase transition upon
heating the insulator. The insulating phase instead smoothly crosses over 
into the high temperature paramagnet via an intermediate cooperative 
paramagnetic regime.
We also study the superfluid-to-normal thermal transition for $V < V_c$. 
We find that
this is a Kosterlitz-Thouless transition at small $V/t$ but
changes to a first order transition for $V$ closer to $V_c$. We argue
that this first order thermal transition is consistent with the 
presence of a nearby $Z_2$ insulating ground state obtained from the 
superfluid ground state by condensing double vortices.
\end{abstract}


\maketitle

%------------------------------------------------------------------------------
\section{Introduction}

Spin liquids are quantum disordered paramagnetic phases that preserve all 
lattice symmetries.
While there has been considerable progress in
understanding the effective field theories and properties of such
spin liquid phases \cite{wen,subir,senthil,motrunich:senthil,kitaev,wen2},
showing 
that the excitations in this phase carry fractional quantum numbers and 
interact with emergent gauge fields, there are very few microscopic
models which can be convincingly shown to exhibit a spin liquid phase
and they fall, roughly, into two categories. 
One class of microscopic Hamiltonians are quantum dimer models
\cite{moessner, misguich}, which have
been proposed to describe spin gapped phases of quantum magnets. Some
of these models, on the triangular and kagome lattice in two dimensions, 
exhibit $Z_2$ fractionalized quantum disordered phases. However, the Hilbert 
space of such quantum dimer models has a strong local constraint, namely,
the number of dimers emerging from a site is fixed. It is therefore
interesting to examine other models which do not have such local Hilbert
space constraints and, thus, no extra conservation laws other than
the total spin or total $S_z$. Under this second category are models, which 
incorporate a so-called ``cluster charging'' energy \cite{senthil:motrunich},
which penalizes spin 
configurations where $S_z$ summed over a local ``cluster'' of sites differs 
significantly from its mean value. 
Some simple models of this type can be 
shown to reduce to effective quantum dimer Hamiltonians in the limit of a 
large charging energy, with the
Hilbert space constraints emerging at low energy from
energetic considerations. By the usual mapping between $S=1/2$ spins and 
hard core bosons, such quantum spin models can be alternatively viewed as
boson models. The spin liquid phase of the spin model at zero magnetization
then corresponds to a uniform Mott insulator of bosons at half-filling.

%---------------------------------------------------------------------------
\begin{figure}[t]
\includegraphics[width=3.0in]{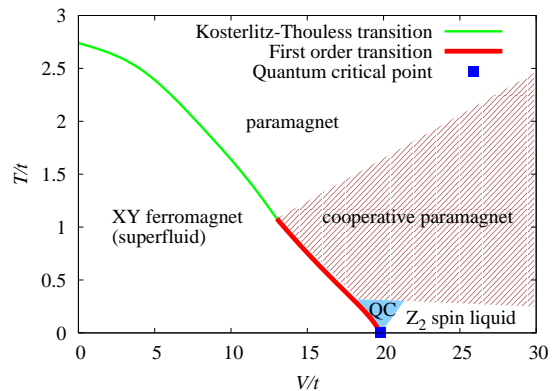}
\caption{ (color online). 
The schematic phase diagram of the model given by Eq.(1) or
equivalently Eq.(2). The superfluid-insulator transition 
in the model (1) corresponds to the XY ferromagnet to spin liquid
transition in the model (2). Notice that the Z$_2$ fractionalized
Mott insulator (Z$_2$ spin liquid) exists only at zero temperature
in two dimensions and as a consequence, there is no finite
temperature phase transition in the paramagnetic region, but
only the crossover between different regimes as explained
in the text. Near the quantum critical point between the
superfluid and the Z$_2$ fractionalized insulator, there should
be a quantum critical region (denoted as QC in the figure) that
is not discussed in the present paper.
}
\label{fig:phasediagram}
\end{figure}
%---------------------------------------------------------------------------

In this paper we study, using a generalized
stochastic series expansion QMC algorithm \cite{sse,footnote.sseplaq}, a
``cluster charging'' model of hard core bosons on the kagome lattice with 
the Hamiltonian
\beq
  H_{\text{b}}=
    -t \sum_{(i,j)}
      \left( b^{\dag}_i b_j + \text{H.c.} \right)
    +V \sum_{\hexagon} (n_{\hexagon})^2
    - \mu \sum_{i} n_i.
\label{eq:bosonmod}
\eeq
Here $b^{\dag}_i$($b_j$) is the boson creation(annihilation) operator,
$t > 0$ is the hopping amplitude, $V > 0$ is the
repulsion strength, $n_i=b^{\dag}_i b_i$ is the number operator, and
$\mu = 12 V$ is the chemical potential which fixes the boson density
to be at half-filling. The hopping term connects only the
first, second and third neighbors. The repulsive interaction is
similarly short-ranged. The main result of this paper is
the phase diagram summarized in Fig.~(\ref{fig:phasediagram}).

We begin by reviewing our earlier results \cite{shortpaper} which show that
model (\ref{eq:bosonmod}) exhibits, at $T=0$, a superfluid-insulator quantum
phase transition at
$(V/t)_c \approx 19.8$. The insulating phase is a uniform,
topologically ordered, $Z_2$ Mott insulator. We then present new results 
for the finite temperature phase diagram. We find that the insulator
crosses over into the high temperature normal phase via a
cooperative fluctuation regime but with apparently no intervening thermal 
phase transitions. This bolsters the case for no broken lattice symmetries 
in the insulating phase. On the superfluid side, for small values of
$V/t$, raising temperature leads to a Kosterlitz-Thouless
(KT) transition from the superfluid phase to a normal phase. As we approach
the quantum critical point however, with $V/t \gtrsim 13$, the KT
transition converts into a first order transition. We argue, from a 
renormalization group analysis of an appropriate sine Gordon model, that
this is consistent with an underlying quantum phase transition into a
$Z_2$ insulating phase driven by double vortex condensation.

Our model in Eq.(\ref{eq:bosonmod}) is inspired by an XXZ spin model for
$S=1/2$ quantum spins proposed by
Balents, Fisher and Girvin \cite{balents1} (BFG), with a Hamiltonian
\beq
  H_{\text{XXZ}}=
    - J_{\perp} \sum_{\hexagon} [(S^x_{\hexagon})^2+(S^y_{\hexagon})^2-3]
    +J_z \sum_{\hexagon} (S^z_{\hexagon})^2.
\label{eq:xxzmod}
\eeq
Here $S^a_{\hexagon}=\sum_{i\in\hexagon}S^a_i$ is a sum over the
six spins on each hexagon of the kagome lattice unit cell,
$\sum_{\hexagon}$ denotes a sum over all hexagons on the lattice.
This model is easily seen to be a
short-range anisotropic XXZ model, with only the first, second and third 
neighbor interactions being nonzero and equal to each other. 

Analyzing the model in Eq.(\ref{eq:xxzmod}) for $J_\perp<0$ and
$J_z/|J_\perp| \gg 1$, and interpreting each up-spin as a dimer on
a triangular lattice, BFG showed \cite{balents1}
that the Hamiltonian is dual to an 
effective triangular lattice quantum dimer model with three dimers touching 
each site. In spin language, this effective model takes the form of a
ring-exchange model, with an exchange scale $J_{\text{ring}}=J^2_\perp/J_z$,
which describes quantum dynamics in the Hilbert space with 
$S^z_{\hexagon}=0$ on each hexagon, the local
constraint arising from energetic considerations at large $J_z/J_\perp$. 
Supplementing this
model with an additional four-site (Rokhsar-Kivelson (RK) \cite{rk})
potential term of strength $v_{4}$ they showed that this modified Hamiltonian
is in spin liquid phase for $v_4=J_{\text{ring}}$, which was argued to be 
stable for small 
deviations $v_4<J_{\text{ring}}$. Later exact diagonalization (ED) numerics 
\cite{balents2} showed that the ring-exchange model appears to be in this 
spin-liquid phase down to $v_4=0$, but only system sizes upto 20 unit cells
could be explored.

The relation
of the XXZ model to the hard core boson model we study follows upon using
the standard mapping between $S=1/2$ quantum spins and hard core bosons.
Specifically, the Hamiltonian we study in this paper at half-filling
is equivalent to that 
in Eq.~(\ref{eq:xxzmod}) if we set $J_\perp = t > 0$ and $J_z = V > 0$.
Since the 
ring-exchange physics is independent of the sign of $J_\perp$, we expect to 
recover, for large values of $J_z$, the physics of the XXZ model with 
$J_\perp < 0$ studied by BFG.  On the technical side, the 
choice of $J_\perp>0$ eliminates the QMC sign problem and allows us to
go significantly beyond earlier work on this class of models.

This paper is organized as follows. Section II reviews some of our earlier 
results and presents some new results on the zero temperature phase diagram,
including a discussion of topological order in the insulator. Section III
discusses the finite temperature region of the insulating phase, including
the temperature dependence of the energy and an estimate of the vison gap.
Section III discusses numerical and analytical results for the finite 
temperature superfluid-normal phase transition. Section IV presents a
summary of the results.

%------------------------------------------------------------------------------
\section{Zero temperature phase diagram}

We begin by reviewing the zero temperature phase diagram of model 
(\ref{eq:bosonmod}) which was studied by us in an earlier paper
\cite{shortpaper}.
For $V/t=0$, the bosons only experience the hard-core constraint, and
therefore condense
into a superfluid phase. As we turn on interactions, the local charging energy $V$
penalizes those configurations where the total number of bosons on any hexagon
deviates from its mean value of $\bar{n}_{\hexagon}=3$. At large $V/t$, this cluster charging 
energy leads to locally incompressible hexagons. This suppresses off-diagonal long 
range order (and superfluidity) and drives the system into an insulating phase.
In the following subsections we review our earlier numerical results which show that 
the superfluid insulator transition is a continuous quantum phase transition with 
$z=1$. We also review our results and present new numerical data which show that the
insulator has four degenerate, topologically distinct, ground states.

\subsection{Quantum superfluid-insulator phase transition}

%---------------------------------------------------------------------------
%\begin{figure}[t]
%\includegraphics[width=3.0in]{rhos_b=18.eps}
%\caption{ (color online). The superfluid density $\rho_s$ versus
%$V/t$, at $T=t/9$ for different linear system sizes $L$. 
%Finite-size effects, significant only near the quantum phase transition,
%are discussed in the text and Fig.~\ref{fig:rhos:collapse:scaling}.
%}
%\label{fig:rhos}
%\end{figure}
%---------------------------------------------------------------------------

For small values of $V/t$, we expect the ground state of model 
(\ref{eq:bosonmod}) to
be a superfluid. We characterize this superfluid phase 
by its superfluid density $\rho_s$ measured through winding number
fluctuations \cite{windingnumber} $W_{a_{1,2}}$ in each of the two lattice
directions, with
$$
\rho_s={\langle W_{a_1}^2\rangle + \langle W_{a_2}^2 \rangle \over 2\beta t},
$$
where $\beta$ is the inverse temperature.
For small $V/t$, $\rho_s$ is large and its value agrees with the mean field
estimate obtained by Gutzwiller projecting a free Bose condensate so as to
satisfy the hard core constraint \cite{unpub}. 
$\rho_s$ decreases with
increasing $V/t$, eventually vanishing for $V/t\gtrsim 20$ suggesting a phase
transition to an insulating phase ($\cI^*$). This behavior is
in sharp contrast to a nearly identical kagome lattice model where the hopping 
and repulsive
interactions are restricted to the nearest neighbor only --- in that case,
\cite{kagome:nn} a uniform superfluid persists for arbitrarily large $V/t$.
The absence of a jump in $\rho_s$ on going through the transition
suggests that the $\cS-\cI^*$ transition is continuous.

%---------------------------------------------------------------------------
\begin{figure}[t]
\includegraphics[width=3.0in]{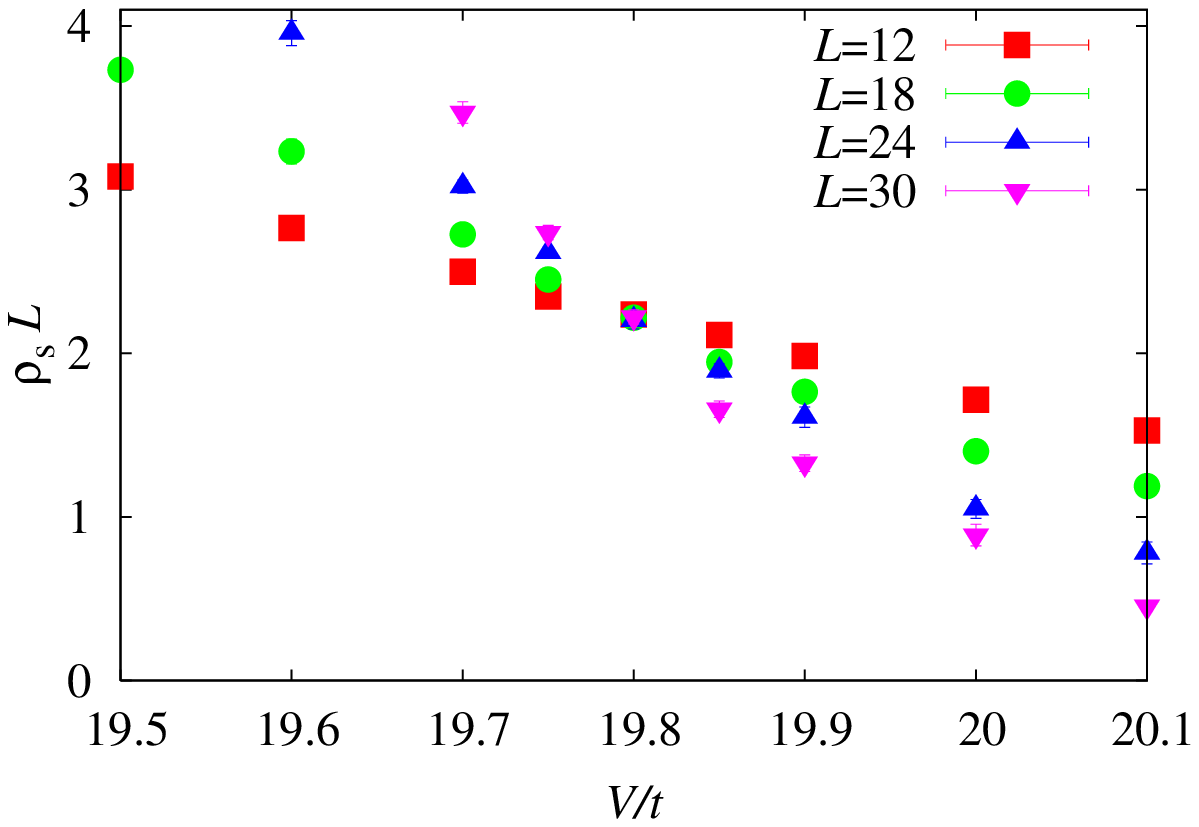}
\includegraphics[width=3.0in]{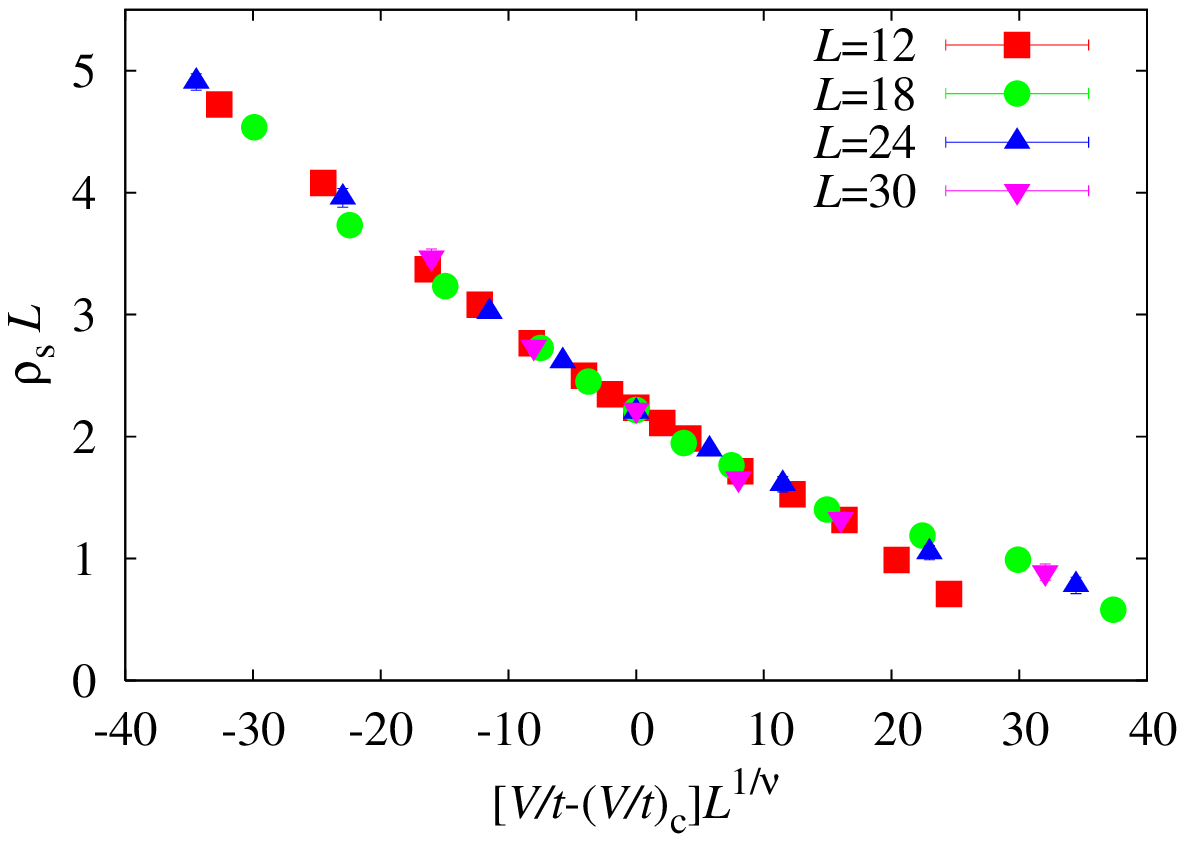}
\caption{ (color online).
Upper panel: Finite size scaling of $\rho_s$ for $\beta/L=3/(4t)$.
Lower panel: Data collapse of the superfluid density $\rho_s$ for
$\beta/L=3/(4t)$, $(V/t)_c=19.80(2)$, and $\nu=0.67(5)$.
}
\label{fig:rhos:collapse:scaling}
\vskip -0.2cm
\end{figure}
%---------------------------------------------------------------------------

In the vicinity of a continuous QPT, the superfluid density should scale as
\beq
  \rho_s=L^{-z} F_{\rho_s}(L^{1/\nu}(K_c-K), \beta/L^z),
\label{eq:scalingrhos}
\eeq
where $F_{\rho_s}$ is the scaling function, $L$ is the linear system size,
$z$ the dynamical critical exponent, $\nu$ the correlation length exponent,
and $(K_c-K) \propto (V_c-V) /t$ is the distance to the critical point.
Thus if we
plot $\rho_s L^z$ as a function of $V/t$ at fixed aspect ratio $\beta/L^z$,
the curves for different system sizes should intersect at the critical point.
The inset of Fig.~\ref{fig:rhos:collapse:scaling} shows such a plot for
$z=1$ and $\beta/L=3/(4t)$ with a clear crossing point
at $(V/t)_c \approx 19.8$. The data is thus consistent with a continuous 
$\cS-\cI^*$ transition with the dynamical exponent $z=1$.
To obtain the correlation length exponent $\nu$, we plot $\rho_s L$ as a
function of $[(V/t)_c-V/t]L^{1/\nu}$ for different system sizes.
It follows from Eq.~(\ref{eq:scalingrhos}) that the curves for different
system sizes should collapse onto a universal curve $F_{\rho_s}$ for
a properly chosen $(V/t)_c$ and $\nu$. In
Fig.~\ref{fig:rhos:collapse:scaling}, we show such a data collapse for
$(V/t)_c=19.80(2)$ and $\nu=0.67(5)$. The error bars are estimated from the
stability of the collapse towards varying the parameters. 
To summarize, we find a continuous $\cS-\cI^*$ transition with exponents
$z=1$ and $\nu=0.67(5)$. We next examine the nature of the insulator $\cI^*$.

%------------------------------------------------------------------------------
\subsection{Characterizing $\cI^*$: Topological degeneracy and
absence of broken symmetries}

In order to test whether the insulating phase of this model is a conventional
broken symmetry state, we have studied correlation functions in $\cI^*$.
We have looked for signatures of diagonal (density), bond or plaquette
ordering by studying the equal-time density and bond structure factors
to check for different ordering
patterns. Even for system sizes as large as $48\times 48$ kagome unit cells
and temperatures as low as $T/J_{\text{ring}}\approx 0.2$, where
$J_{\text{ring}}=t^2/V$,
%$T/J_\perp\approx 0.01$,
we find no evidence of any Bragg peaks, or any ordering tendency,
in these correlators. This appears
to rule out the possibility that $\cI^*$ is a conventional lattice symmetry 
broken state. Additional evidence for a uniform insulating phase
comes from the fact that if the insulator had broken lattice symmetries
it would not be smoothly connected to the high temperature paramagnet but
be necessarily be separated from it by a thermal phase transition.
However, we find no apparent signs of any thermal phase transition upon 
heating up from $T=0$ towards the uncorrelated high $T$ paramagnet.

For a system of bosons,
momentum counting arguments \cite{oshikawa, ashvin} show
that an insulating state at half-filling could
either be a conventional state with broken lattice symmetries or must
{\em necessarily} have topological order which means the
ground state degeneracy depends on the topology of the system. Since
we have ruled out, as best as we can, the possibility that the insulating
phase breaks lattice symmetries, we next examine the insulating phase for
signs of topological order.

%------------------------------------------------------------------------------
\begin{figure}[t]
\includegraphics[width=3.0in]{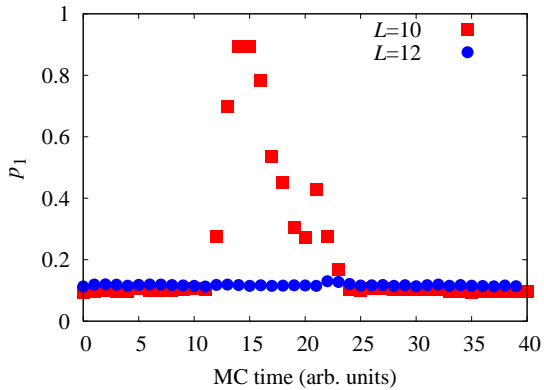}
\caption{ (color online).
Average parities as a function of Monte Carlo time for two different system
sizes and $T=t/20$ in the insulator at $V/t = 24$. The start configuration
has $p_{1,2}=0$. The parities are finite because there is a small density of
defects. The parities for $L=12$ do not fluctuate indicating that the system
does not change its topological sector whereas they strongly fluctuate for
$L=10$ indicating that the topological sector is changing.
}
\label{fig:parities}
\end{figure}

On a lattice with periodic
boundary conditions in both lattice directions,
the subspace of configurations with $n_{\hexagon}=3$ on every hexagon of
the kagome lattice is
identical to the Hilbert space of a triangular lattice quantum dimer model,
with three dimers touching each site (identifying the hardcore boson 
with a dimer).
It is well known that such quantum dimer models have well defined topological
sectors, distinguished by the eigenvalues of 
a nonlocal operator, which do not mix under
the dynamics generated by the Hamiltonian. In the context of our model, if
we set $t/V=0$ and examine the classical ground states, which do
respect $n_{\hexagon}=3$, these
sectors correspond to having, for each lattice direction
$a_{1,2}$, an odd (or even) number of bosons on each row (or column) of the
lattice
(so called ``parity sectors'').
The row/column parities defined in this manner {\it do not}, however, specify
topological sectors of
model (\ref{eq:bosonmod}) since for any nonzero $t$, no matter how small, 
there will be a 
small density of defect hexagons \cite{footnote.defects} with $n_{\hexagon} 
\neq 3$ so that the row or column
parity is not conserved under the Hamiltonian dynamics.
How do we then look for topological sectors in the ground state
of model (\ref{eq:bosonmod})?

We have checked in our QMC numerics, that if we start from a configuration 
which lies in the dimer subspace with a certain row/column parity, then
the QMC algorithm
generates a small density of particle-hole pair defects in equilibrium,
so that the quantum ground state no longer lies in the ``dimer subspace''. 
However, for a large enough linear
system size $L$ at a given value of $V/t$ (e.g., $L \gtrsim 10$ at $V/t=26$),
our simulations with the longest accessible MC steps do not 
lead to {\it any} non-local boson moves which wind around the lattice, 
see Fig.~\ref{fig:parities}. Thus the winding number 
identically vanishes, and each configuration in the simulation which
lies in the dimer subspace belongs to the same parity sector as the
initial configuration. 
The full ground state accessed by the QMC is, in this sense,
perturbatively connected to the initial parity sector. In other words,
we can start with the equilibrium QMC ground state and erase nearby
particle-hole defects in pairs and obtain a state which lies entirely in 
the starting topological sector. It is in this sense that
we can identify the four topological sectors even for model 
(\ref{eq:bosonmod}), and we can continue to label 
them simply by the row/column parity of that component of the ground 
state wavefunction which lies in the dimer subspace.
The four ground state wavefunctions can be thus be written as 
\bea
  &&|\psi_{00} \rangle = |\psi_{00}^{d} \rangle + |\psi'_{00}\rangle, \nonumber\\
  &&|\psi_{01} \rangle = |\psi_{01}^{d} \rangle + |\psi'_{01}\rangle, \nonumber\\
  &&|\psi_{10} \rangle = |\psi_{10}^{d} \rangle + |\psi'_{10}\rangle, \nonumber\\
  &&|\psi_{11} \rangle = |\psi_{11}^{d} \rangle + |\psi'_{11}\rangle, \nonumber
\eea
where $|\psi_{ab}^{d} \rangle$ denote the components of the wavefunctions
that lie in the dimer subspace defined by parities $a$ and $b$ and
$|\psi'_{ab}\rangle$ denote the components of the wavefunctions that
do not lie in the dimer subspace but are connected to $|\psi_{ab}\rangle$
by short range hops of the bosons. It is clear that these wavefunctions are
distinct eigenstates of our local Hamiltonian and they are not
connected by local moves.
For a {\em topologically ordered} insulator, the ground state energy should 
be {\em identical} in each of the four topological sectors (on a torus) 
leading to a ground state degeneracy of four. We have computed the 
energy of the four ground states by starting our simulations from 
configurations in the dimer subspace lying in four different parity sectors.
We find that they are equal within statistical errors, which is strong 
evidence for topological order \cite{footnote.toporder}. 

%------------------------------------------------------------------------------
\subsection{Vison correlations}

%---------------------------------------------------------------------------
%\begin{figure}[t]
%\includegraphics[width=3.0in]{vcorr2_l=24_jz=20.5_b=36.eps}
%\caption{ (color online).
%Equal-time vison-vison correlation function for $L=\!24$ and $T=t/18$ in
%the insulator at $V/t = 20.5$, showing exponential decay with a
%length scale $\xi = 1.43(5)$.
%}
%\label{fig:visoncorr}
%\end{figure}
%---------------------------------------------------------------------------

A second signature of
$Z_2$ fractionalization is that visons, which are
gapped $Z_2$ vortices in the effective field theory description, should
have exponentially decaying spatial correlations.
The spatial vison-vison correlation function is the expectation
value of a string operator in terms of the spins. For the ring-exchange
model (valid for $V/t \to \infty$), it takes the form \cite{balents1}:
\beq
  C_{\text {vv}}(r_{ij}) = |\langle 0| 
      \prod_{k=i}^{j}\!\!\!\!\!\!\!\!\!\!\longrightarrow  e^{i\pi n_k}
%	   \prod_{k=i}^{j} e^{i\pi n_k}
    |0\rangle|,
\label{eq:vcorr}
\eeq
where $|0\rangle$ denotes the ground state and the product is along some
path on the kagome lattice that contains an even number of sites, starts at
site $i$, and ends at site $j$, making only ``$\pm 60^\circ$'' turns to
the left or right. $n_k=0,1$ is the number of bosons at site $k$.
The absolute value of the product in Eq.~(\ref{eq:vcorr})
is {\em path-independent} in the dimer subspace,
and it is expected to decay exponentially in the topologically 
ordered phase. 
In model (\ref{eq:bosonmod}) at finite $V/t$, ground state no longer lies 
entirely in the dimer subspace, but will mix in configurations with 
particle-hole defects. However, in the same manner as we have used
the dimer subspace component of the wavefunction to define topological
sectors, we can similarly use that wavefunction component to also compute 
the vison correlator.
We have found that $C_{\text {vv}}(r_{ij})$, computed
by essentially dropping all configurations containing particle-hole defects,
decays exponentially in $\cI^*$, again signaling topological order in
$\cI^*$. At $V/t=20.5$, we estimate a ``decay length'', $\xi=1.43(5)$, which
is comparable to its value at the RK point of the ring-exchange model
\cite{balents1}, $\xi\approx 1$, and to that found by ED \cite{balents2}, 
$\xi\approx 1.7$, in the ring-exchange model with $v_4=0$. This exponential
decay shows us that significant local particle rearrangements and
fluctuations are possible even in the insulating phase --- such fluctuations
are generated by the effective ring exchange dynamics of the bosons.

%------------------------------------------------------------------------------
\section{Heating the insulator: Cooperative paramagnetic regime
and the vison gap}
\label{seq:visongap}

%---------------------------------------------------------------------------
\begin{figure}[t]
\includegraphics[width=3.0in]{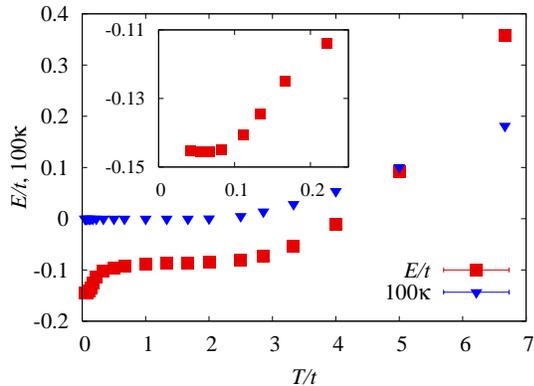}
\caption{ (color online). Energy per site $E$ and compressibility
$\kappa$ versus temperature for $L=24$ ($V/t$=20.5). The energy 
rises exponentially at very low $T$ (see inset) with a wide intermediate 
plateau from $T/t \sim 0.5-3$ (see text). The compressibility is zero 
(within error bar) at low $T$, rising only at $T\sim 2.5 t$ once gapped
charge (spinon) excitations become relevant.}
\label{fig:energy}
\vskip -0.2cm
\end{figure}
%---------------------------------------------------------------------------

To provide further evidence for gapped vison excitations,
we display the temperature dependence of the system energy per site and
compressibility
$$
  \kappa = \frac{\beta}{N}
   \left\langle \left( \sum_i n_i \right)^2 \right\rangle
$$
in Fig.~\ref{fig:energy}.
Upon heating up from the ground state, the energy exhibits a two-step increase, 
with a distinct intermediate plateau. 
At the lowest temperature, the energy increases exponentially from the 
spin-liquid ground state as seen from the inset of Fig.~\ref{fig:energy}. 
The energy then reaches a plateau at a temperature
$T \sim J_{\rm ring}$. This plateau corresponds to a regime where the system 
dominantly explores configurations with $n_{\hexagon}=3$ on each hexagon. In
spin language, it corresponds to a ``cooperative paramagnet''. Upon heating
further, the energy increases beyond its plateau value when the temperature
becomes comparable to the local charge gap set by $V$.
We confirm this by noting that there is a sharp increase in $\kappa$ at 
this higher temperature (also shown in Fig.~\ref{fig:energy}). 

%The energy exhibits a two-step decrease 
%upon lowering temperature, with a distinct intermediate plateau, before
%vanishing exponentially at a very low temperature. We identify the first
%drop in energy with a freezing out of charge fluctuations below a charge
%gap scale, which we confirm by the sharp decrease in $\kappa$ at
%this temperature (also shown in Fig.~\ref{fig:energy}). The plateau then
%corresponds to a regime where the system dominantly explores
%configurations with $n_{\hexagon}=3$ on each hexagon. In spin language,
%it corresponds to a ``cooperative paramagnet''.
%Finally at the lowest temperature, the system begins evolving into the 
%spin-liquid ground state. The absence of any clear signs of a phase
%transition between the insulating ground state and the high temperature
%phase supports the absence of any broken (discrete) lattice symmetries
%in the insulator.

Heating up from $T=0$, we therefore identify the 
lowest energy excitations out of the ground state as coming from
vison-pair excitations of the spin liquid (since the charge gap is much larger). 
The temperature dependence 
of the energy thus gives us a rough idea of the single vison gap; for
$V/t=20.5$, we estimate $E_v/t \sim 0.35(15)$.

Note that the classical model with $t=0$ has a large entropy at $T=0$ arising 
from a macroscopic number of degenerate classical ground states. When one
turns on a nonzero $t$, this degeneracy is lifted as the ground state becomes
a quantum spin liquid which supports vison excitations ($Z_2$ vortices). From the 
point of view of the spin liquid ground state, we can therefore view
the large entropy density of the classical dimer state as arising from the 
large entropy density of multiple vison excitations as the system is heated.
This means that the energy curve should have a finite temperature plateau at
the energy level with
the largest entropy density where, very roughly, half of all allowed visons get
excited which contributes to a large configurational entropy of visons. This
happens at an energy $E_s\approx E_0/2$ as measured from
zero classical ground state energy, where $E_0$ is the quantum ground state
energy. This is consistent with our numerical data.
 
%------------------------------------------------------------------------------
\section{Heating the superfluid: Superfluid-normal phase transition}

For $V < V_c$, the system is in a superfluid ground state. Heating this 
superfluid leads to transition from a superfluid phase to a normal liquid
phase at finite temperatures. This transition is usually of
a Kosterlitz-Thouless (KT) type \cite{berenzinskii, kt} that is driven by
vortex unbinding. In principle,
this transition can also be first order when the vortex core energy is
small enough \cite{caillol, minnhagen}. 
We find from our numerics that the thermal superfluid normal
transition is a KT transition at small enough values of $V/t$. This
goes along with the conventional wisdom. However, the transition becomes
first order roughly at $V/t \gtrsim 13(1)$. This appears to be a
novel example of a discontinuous thermal superfluid-normal transition in 
a microscopic two-dimensional quantum model. 

\subsection{QMC results for the SF-normal thermal transition}
\begin{figure}[t]
\includegraphics[width=3.0in]{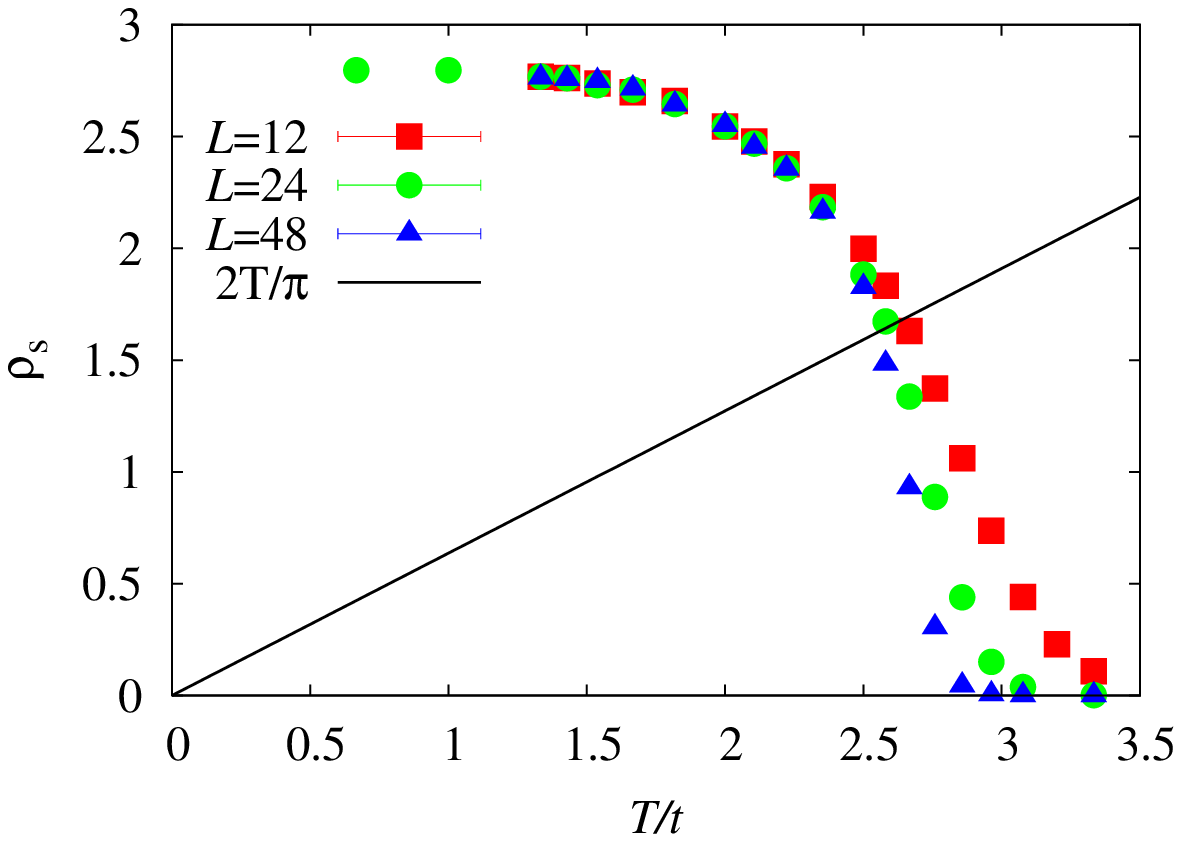}
\includegraphics[width=3.0in]{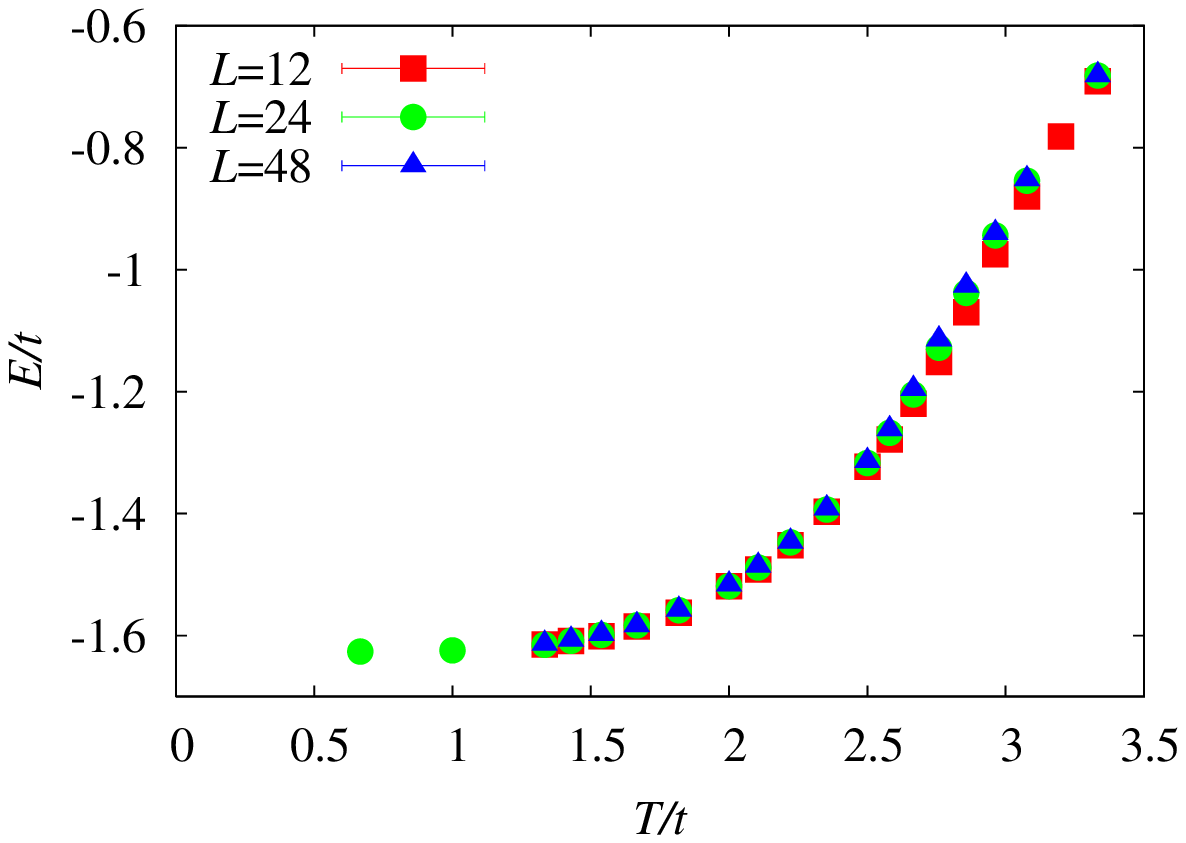}
\caption{ (color online). The superfluid density $\rho_s$ and the energy $E$
as a function of temperature at $V/t=16$.
}
\label{fig:kt}
\end{figure}

In Fig.~\ref{fig:kt}, we show the superfluid density $\rho_s$ and the
system energy $E$ as a function of temperature at $V/t=4$. Both quantities
exhibit smooth behavior. The superfluid density should be discontinuous at
the KT transition with the universal jump
$$
  \Delta \rho_s = {2T_{\text{KT}} \over \pi},
$$
where $T_{\text{KT}}$ is the KT critical temperature. However, this
discontinuity is approached only logarithmically as the system size
increases. The RG equations \cite{kosterlitz} predict that in the vicinity
of the KT transition the superfluid density scales as
\cite{weber:minnhagen,olsson,harada:kawashima}
\beq
   \rho_s=\frac{2T}{\pi} \left\{1+
      \frac{F[(T-T_{\text{KT}})\ln^2(L/L_0)]}{2\ln(L/L_0)}\right\},
\label{eq:ktscaling}
\eeq
where $F$ is the scaling functions such that $F(x)\approx 1+O(x)$ for small
values of $x$, $L$ is the linear system size, and $L_0$ is some length scale.
If one plots $(\pi\rho_s/T-2)\ln(L/L_0)$ as a function of
$(T-T_{\text{KT}})\ln^2(L/L_0)$
then it follows from Eq.~\ref{eq:ktscaling} that the curves for different
system sizes should collapse onto a universal curve $F$ for a properly
chosen $T_{\text{KT}}$ and $L_0$. In addition, $(\pi\rho_s/T-2)\ln(L/L_0)$
should take a value of $1$ at the KT transition point.
%---------------------------------------------------------------------------
%---------------------------------------------------------------------------

%---------------------------------------------------------------------------
\begin{figure}[t]
\includegraphics[width=3.0in]{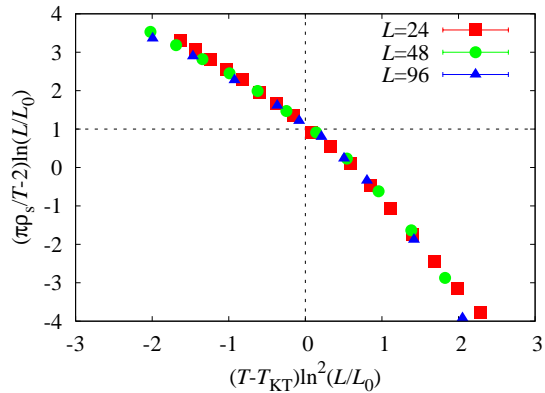}
\caption{ (color online).
Data collapse of the superfluid density $\rho_s$ for $T_{\text{KT}}=2.486t$
and $L_0=2$. Horizontal and vertical dashes lines are guides for the eye.
}
\label{fig:ktcollpase}
\end{figure}
%---------------------------------------------------------------------------

In Fig.~\ref{fig:ktcollpase}, we show such a data collapse for
$T_{\text{KT}}=2.486t$ and $L_0=2$. Thus we can conclude that the transition
at $V/t=4$ is a KT transition.

%---------------------------------------------------------------------------
\begin{figure}[t]
\includegraphics[width=3.0in]{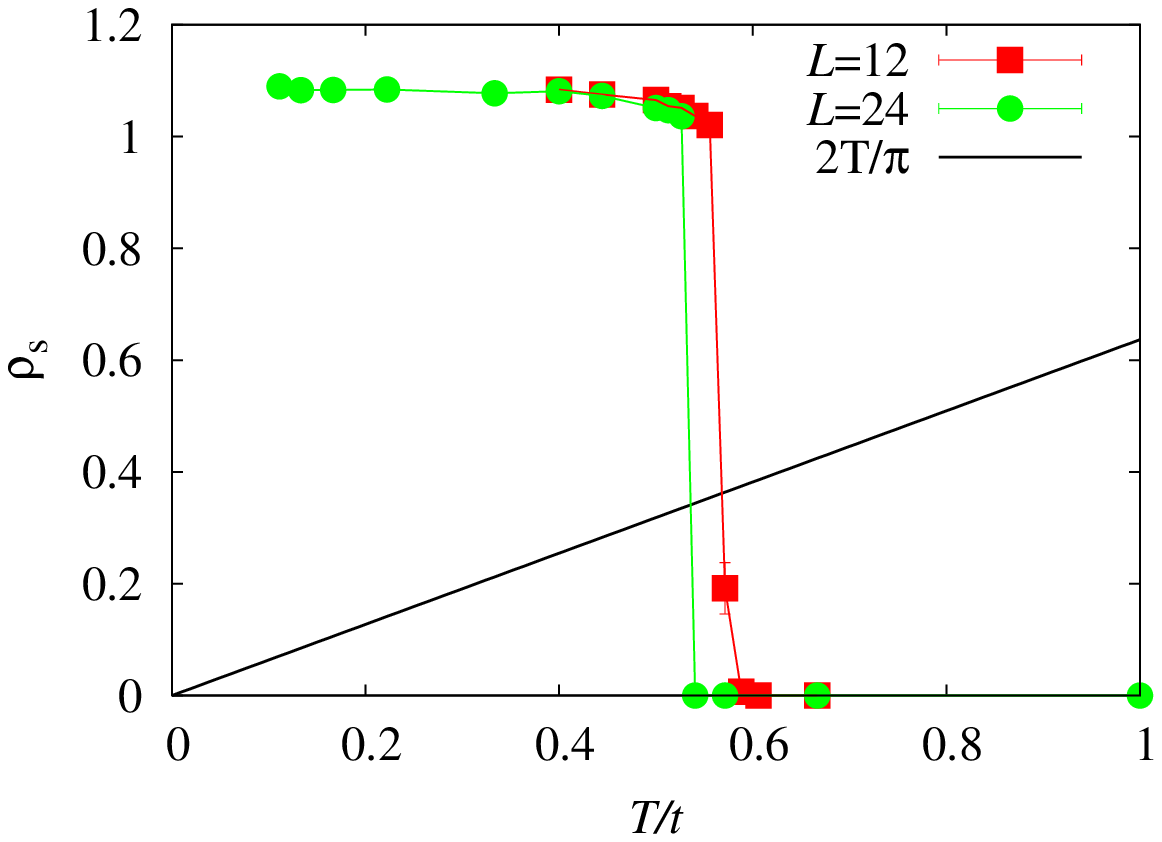}
\includegraphics[width=3.0in]{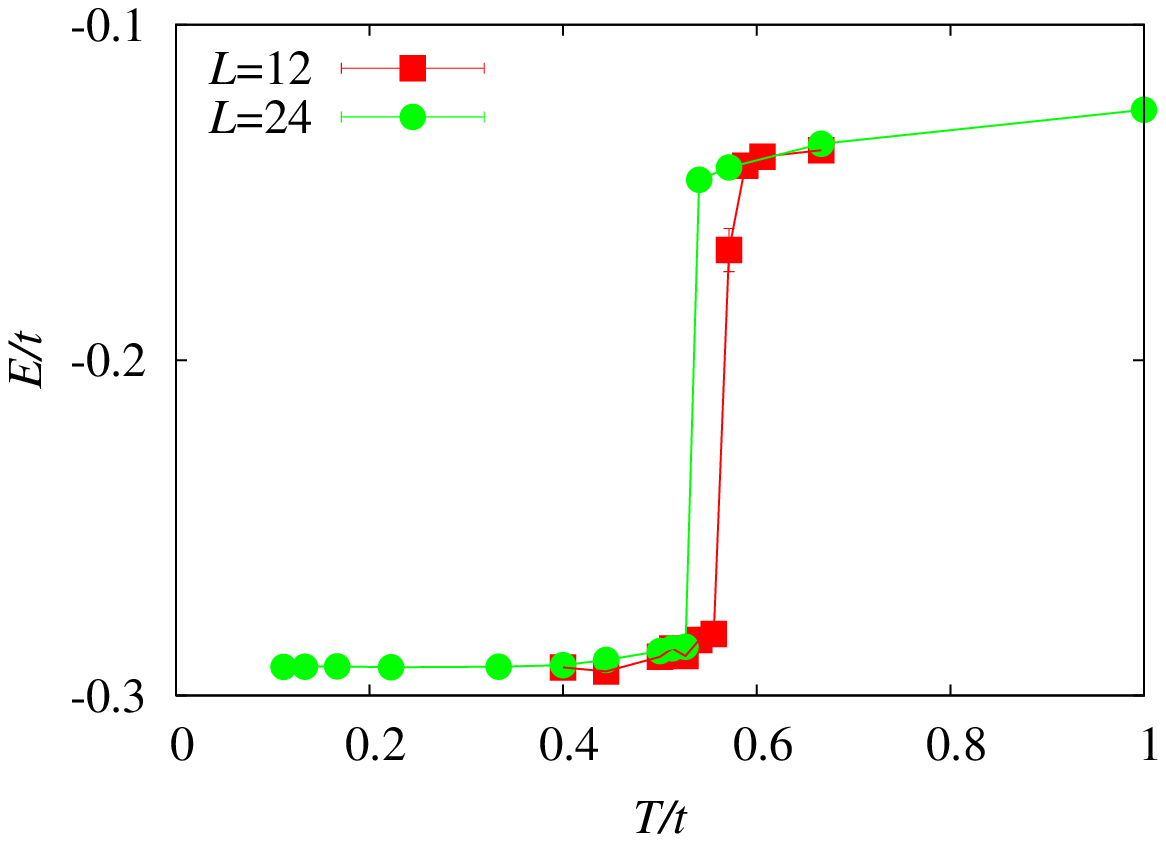}
\caption{ (color online). The superfluid density $\rho_s$ and the energy $E$
as a function of temperature at $V/t=16$.
}
\label{fig:1storder}
\end{figure}
%---------------------------------------------------------------------------

%---------------------------------------------------------------------------
\begin{figure}[t]
\includegraphics[width=3.0in]{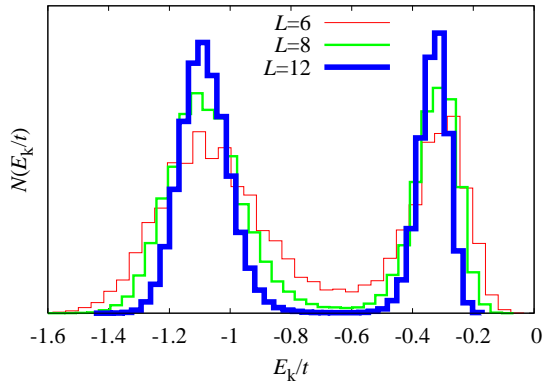}
\caption{ (color online).
Distribution of the kinetic energy close to the transition for different
systems sizes and slightly different temperatures ($\beta t=1.67$ for $L=6$,
$\beta t=1.745$ for $L=8$, and $\beta t=1.77$ for $L=12$) at $V/t=16$.
A double peak structure is visible even for very small systems sizes and
it becomes more pronounced, leading to two well-separated peaks,
as the system size increases indicating a strongly
first order transition.
}
\label{fig:histogram}
\end{figure}
%---------------------------------------------------------------------------

The situation is strikingly different for values of $V/t$ roughly
larger than $13$.
In Fig.~\ref{fig:1storder}, we show the superfluid density and the system
energy as a function of temperature at $V/t=16$. Both the superfluid
density and the energy jump suggesting that the transition is first order.
As can be seen form Fig.~\ref{fig:histogram}, the distribution of the
kinetic energy, $E_{k}=-t\langle (b^{\dag}_i b_j + \text{H.c.})\rangle$,
close to the transition has a clearly visible double peaked structure even
for very small systems sizes indicating that the transition is strongly first
order. The double peaked structure becomes more pronounced, leading to
two well-separated peaks, as the systems
size increases. We also observe hysteresis effects by crossing the
transition point upon heating or cooling the system (not shown). This is
also indicates that the transition is first order.

For
$13 \lesssim V/t \lesssim 17.5$, the normal liquid just above $T_c$ has an
energy which is nearly temperature independent for a range of temperatures.
In this sense, it is analogous to the finite temperature plateau regime
shown in Fig.~\ref{fig:energy} which is obtained upon heating the insulator.
This normal liquid
can be well approximated by a classical liquid with $n_{\hexagon}=3$ on
each kagome hexagon, a so-called ``cooperative paramagnet'' in spin language. 
However, when we increase $V/t$ in the range $17.5 \lesssim V/t \lesssim V_c/t$, 
the normal liquid just above $T_c$ has increasingly significant quantum 
dynamics (as it lies in the quantum critical regime of the quantum phase
transition) and, consequently, a progressively lower entropy than the classical 
liquid with $n_{\hexagon}=3$ on
each kagome hexagon. The first order thermal transition correspondingly gets 
weaker for $V \to V_c$ and eventually merges into the continuous quantum 
phase transition at $V_c$.

%------------------------------------------------------------------------------
\subsection{Analysis of the thermal superfluid-normal transition}

We next turn to an analysis of the thermal superfluid-normal transition
in order to shed some light on the observation that it becomes first 
order rather being a continuous Kosterlitz-Thouless transition
when $V/t$ is close to its quantum critical value. One qualitative
way to understand this physics is to appeal to an entropy mismatch
argument. At low $T$ in the superfluid, the only relevant excitation
is the superfluid sound mode. This has a velocity $c_s$ which is nonsingular
at the $z=1$ quantum phase transition into the insulator. The low temperature
specific heat and entropy thus scale as $\sim T^2/c_s$ deep in the 
superfluid. At large $V/t$ the finite $T$ normal state, if it lies in the
``cooperative paramagnet'' regime, is highly constrained 
and can be described very crudely by all possible classical configurations
for which $n_{\hexagon}=3$ on each hexagon of the kagome lattice. This
classical picture where we ignore all quantum fluctuations then predicts
a large constant entropy $S_{\rm cl}$ in the normal state.
%(This large entropy can also be viewed as
%arising from a large number of multi-vortex excitations which derive
%from the multiple vison excitations on the insulating side of the transition.)
Extrapolating 
from the low temperature superfluid state to the normal state thus leads 
to a large entropy 
mismatch, $S_{\rm cl} - T^2_c/c_s$, at the transition when $T_c$ is small. 
Clearly, once $T_c$ decreases below
a certain value, this entropy mismatch cannot be satisfied by the 
configurational entropy of a {\it dilute} gas of vortices. The superfluid to
normal transition must, at this stage, involve producing a large number
of vortices over a small temperature interval to make up for this entropy
mismatch, implying a large vortex fugacity. Such a large vortex
fugacity is well known to modify the KT transition into a first order
transition. Of course this argument does not hold very close
to the quantum critical point where the normal state just above $T_c$
is not a simple classical liquid with an entropy $S_{\rm cl}$, but is
instead a quantum critical liquid.

In what follows, we will provide a slightly different view of the 
thermal transition. It is well known that a $Z_2$ fractionalized
insulating ground state can be obtained from a superfluid ground state
by condensing double vortices instead of single vortices. If the
superfluid is proximate to such an exotic insulating ground state,
as our other results show, then
the thermal excitations of such a superfluid must include low lying
double vortex excitations in addition to single vortices. Our aim
here will be to infer the presence of these low energy double vortices 
from the observed first order superfluid-normal transition.
We will therefore attempt to study the thermal SF-normal transition
taking into account both single and double vortices in a sine Gordon 
model.

\subsection{Phenomenological model}

Our discussion will use the classical sine Gordon model to describe the 
thermal phase transition from the superfluid to the normal phase. To obtain 
this, we begin with the classical XY model written in vortex language
\be
\beta H_v = 2\pi^2 K \sum_{\br\br'} n_\br n_{\br'} G(\br-\br') + \sum_\br 
\beta E_c(n_\br)
\ee
where $n_\br$ is the vortex number, $G(\br-\br')$ is the vortex interaction
(which is logarithmic at large distances), $K$ is the superfluid stiffness
normalized by the temperature,
and $E_c(n_\br)$ is the local core energy of a vortex with vorticity $n_\br$.
We expect that at a fixed temperature $T$, the normalized stiffness $K$
will decrease as we increase $V/t$. The hope is that such a classical
description is adequate to qualitatively capture the physics of the
thermal transition with quantum fluctuations being important in fixing
the parameters of this classical vortex Hamiltonian.

We can go to $\bk$-space where the vortex interaction takes the simple
form $G(\bk) = 1/(4-2\cos k_x - 2\cos k_y)$ for a 2D square lattice. 
Actually, the only important thing is that $G(\bk \to 0) = 1/\bk^2$.
The detailed form and the lattice geometry are unimportant.
We can then do a Hubbard-Stratonovitch decoupling of the vortex 
interaction term, writing
\bea
&&{\rm e}^{- \sum_\bk 2\pi^2 K n_\bk n_{-\bk} G(\bk)} \\
&\sim& \int D\phi_\bk\phi^*_\bk {\rm e}^{-\sum_\bk \frac{|\phi_\bk|^2}{
8\pi^2 K G(\bk)} + i \sum_\bk \phi_\bk n_{-\bk}}
\eea
so that the partition function takes the form
\bea
Z &\sim& \int D\phi_\br \sum_{\left\{n_\br\right\}} 
{\rm e}^{-\sum_{\la\br\br'\ra} (\phi_\br-\phi_{\br'})^2/(8\pi^2 K)
+ i \sum_\br \phi_\br n_\br} \\
&\times& 
{\rm e}^{-\sum_\br \beta E_c(n_\br)}
\eea
Let us assign core energies $\beta E_c(0)=0$, $\beta E_c(1)=e_1$, 
$\beta E_c(2)=e_2$
and $\beta E_c(n > 2)=\infty$ to simplify the situation. In that case, 
doing the sum over $n_\br$ at each site leads to
\bea
Z &\sim& \int D\phi_\br
{\rm e}^{-\sum_{\la\br\br'\ra} (\phi_\br-\phi_{\br'})^2/(8\pi^2 K)}\\
&\times& \prod_{\br} (1 + v_1 \cos\phi_\br + v_2 \cos2\phi_\br).
\eea
where $v_p = 2{\rm e}^{-e_p}$.
For small $v_p$, we can re-exponentiate the cosine terms to get
\bea
Z &\sim& \int D\phi_\br
{\rm e}^{-\sum_{\la\br\br'\ra} (\phi_\br-\phi_{\br'})^2/(8\pi^2 K)}\\
&\times&
{\rm e}^{\sum_\br (v_1 \cos\phi_\br + v_2 \cos 2 \phi_\br)}.
\eea
Relabelling $1/(8\pi^2 K) \to g/2$, and in the continuum,
\be
Z_{SG} = \int D\phi_\br
{\rm e}^{- \int d^2\br \left[ \frac{g}{2} (\nabla \phi)^2 
 - v_1 \cos\phi - v_2 \cos 2 \phi \right] }.
\ee
The usual XY model corresponds to setting $e_2 = \infty$, or $v_2 = 0$.
Let us refer to these sine-Gordon models as $SG_2$ (with $v_2
\neq 0$, containing
double vortices in addition to single vortices) and 
$SG_1$ (with $v_2=0$, with just single vortices).
We can now study the RG flow equations for $SG_2$, looking at the
combined flow of $g, v_1, v_2$. 
In terms of dimensionful quantities, which will be of use later
while trying to understand the numerical phase diagram, we have
$K = \rho_s/T$ ($\rho_s$ being the superfluid stiffness), so that
$g = T/(4\pi^2 \rho_s)$ and $v_n = 2\exp[-\beta E_c(n)]$.

\subsection{RG equations}

The RG begins by writing $\phi = \phi_> + \phi_<$, where $\phi_>$ has Fourier
modes with momenta $\Lambda {\rm e}^{-d\ell} < q < \Lambda$, while the
slow field $\phi_<$ has the other Fourier components. $\Lambda$ is the
upper momentum cutoff. We fix it so that $\pi \Lambda^2 = 4\pi^2$, or
$\Lambda = 2\sqrt{\pi}$. Integrating over the fast fields, we arrive at 
the following RG flow equations
\bea
\frac{d v_1}{d\ell} &=& v_1 (2 - \frac{1}{4\pi g}) + A v_1 v_2 
\label{flowv1}\\
\frac{d v_2}{d\ell} &=& v_2 (2 - \frac{1}{\pi g}) - B v_1^2 \\
\frac{d g}{d\ell} &=& C_1 v_1^2 +  C_2 v_2^2
\eea
where the coefficients
$A=\frac{\alpha_{1,1}(g)}{4\pi g}$,
$B=\frac{\alpha_{1,1}(g)}{16\pi g}$,
$C_1 =\frac{\alpha_{3,1}(g)}{128\pi^2 g}$, and
$C_2 =\frac{\alpha_{3,4}(g)}{8\pi^2 g}$. Here, we have defined
\be
\alpha_{m,n}(g) = \int_1^\infty dx x^{m-\frac{n}{2\pi g}} J_0(x).
\ee
The lower limit of this integral defining $\alpha_{m,n}(g)$
is set to unity --- this reflects the short distance cutoff from the
lattice spacing (which we have set to $\sim 1/\Lambda$) in doing 
various spatial integrals in this calculation.

Physically, the second order correction in the flow equation
for $v_1$ may be viewed as arising from a single antivortex combining 
with a double vortex to give a single vortex. Similarly the second order
correction to $v_2$ can be viewed as arising from a double vortex
splitting into two single vortices.
For positive $\alpha_{1,1}$ (which is 
the case as we find numerically), the effect of a nonzero $v_2 \gg v_1$
is that it enhances $v_1$ due to the second order coupling Eq.~(\ref{flowv1}).
The main observation we make is that
{\it integrating out soft double vortices renormalizes the single vortex 
fugacity to be larger}.

We can simplify the above RG flow equations by working around $g
\sim 1/8\pi$. Over a range of $g$ near this value,
we approximate $\alpha_{1,1}(g) \approx 5.5 g$, 
$\alpha_{3,1}(g) \approx 6.5 g$, and $\alpha_{3,4}(g) \approx 1.5 g$.
Thus, $A \approx 0.45$, $B \approx 0.11$, $C_1 \approx 0.005$,
$C_2 \approx 0.02$. With these simplifications, we have numerically
studied the above RG flow equations.

\subsection{Flows for $v_1 \ll 1$ and $v_2 \ll v_1$}

For initial $v_2(0)=0$ and
$v_1(0)\ll 1$, we recover the Kosterlitz Thouless RG flow. Namely, 
there is a line of fixed points $(v_1,v_2,g) \equiv (0,0,g^*)$ with
$0< g^* < 1/8\pi$. The termination of this line of fixed points is
the KT transition point at which the stiffness (normalized by the
transition temperature), $K^*=1/(4\pi^2 g^*)
= 2/\pi$, is a universal number. Beyond this point, $g$ flows to 
strong coupling, so that the superfluid stiffness $K \to 0$,
signalling the non-superfluid phase.  For a nonzero $v_2(0) \ll v_1(0)$, 
the above picture remains unchanged, i.e., the Kosterlitz-Thouless
transition is unaffected by a small fugacity of double-vortices.

\subsection{Flows for $v_1 \ll 1$ and $v_2 \sim 1 \gg v_1$}

\begin{figure}
\includegraphics[width=3.2in]{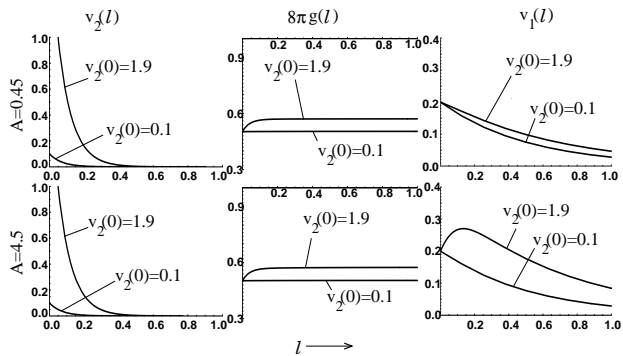}
\caption{Numerically computed flows for $v_1,v_2,g$ shown for two
different choices of $v_2(0)$. We have fixed a small $v_1(0)=0.2$ 
and $8\pi^2 g(0)=0.5$. The top panels shows the flows for $A=0.45$ in
the flow equations which is the result of our perturbative 
calculation. The bottom panels show the flows for an artifically
large $A=4.5$ to indicate that it leads to an enhancement of $v_1$.}
\label{fig:flow}
\end{figure}

For large $v_2(0)$, we find that
$v_2(\ell)$ very quickly renormalizes to small values since it is
strongly irrelevant in the superfluid phase. However, in the initial
stages of the RG flow, it significantly affects the flow of $v_1(\ell)$.
While $v_1(\ell)$ tends to decrease due to the first order term
$(2-1/4\pi g) v_1$, this decrease is partially offset by the positive
contribution from the second order term which couples $v_1$ and $v_2$.
For our calculated couplings, in the regime where
$v_1$ is eventually irrelevant, it flows to zero more slowly for
nonzero $A$.  A comparison of the flows of $v_1$ and $g$ for small and
large $v_2(0)$ is shown in the upper panel of Fig.~\ref{fig:flow}.

A more striking result is obtained by artifically increasing the
coefficient $A$ in the flow equation for $v_1$. 
We find that increasing this to values larger than that given
by the above perturbative calculation leads to a dramatic rise of
$v_1$ in the initial stages of the RG flow, due to coupling with 
$v_2$.  This is illustrated in the lower panel of
Fig.~\ref{fig:flow}, where we have 
chosen $A = 4.5$ instead of our calculated result $A \approx 0.45$.
This modification of the flow equation is in a purely
phenomenological spirit. Such an enhancement might be possible 
once we include the effect of quantum fluctuations or higher winding 
number vortices which we have ignored, but which do become important 
near the superfluid-insulator quantum phase transition; however, this 
is beyond the scope of our perturbative analysis.

Fig.~\ref{fig:flow}
is the central qualitative result of our RG calculation. Namely, 
the interconversion between double vortices and single vortices together
with the low core energy (a large bare fugacity) for double vortex 
excitations can lead to a significantly enhanced fugacity for single 
vortices. At the same time, the double vortices are themselves irrelevant 
at long length scales.

We will next use this strongly enhanced single vortex fugacity,
obtained at intermediate length scales in the RG, to argue for a first 
order superfluid-normal thermal transition. Let us begin from the 
$SG_2$ sine-Gordon theory with $v_2(0) \gg v_1$(0), and follow the RG 
flows until we reach a fixed length scale $\xi$ where $v_2(\xi) \ll v_1(\xi)$.
At this stage, we can drop $v_2$ altogether and study $SG_1$ with 
only $v_1 \neq 0$. We have shown that at some
intermediate scale, $v_1(\xi)$ can become large. In order to accommodate
this large $v_1(\xi)$ within the $SG_1$ theory, the $SG_1$ action must 
be tuned to have a large bare fugacity for single vortices --- in other
words, if we ignore double vortices (which we have shown is reasonable),
the large $v_1(\xi)$ must be viewed as arising from a large $v_1(0)$. 

\subsection{Analysis of the usual sine-Gordon model}

From our above RG analysis, we conclude that it makes sense to capture
the effect of double vortices by studying the effect of a large bare $v_1$ 
in the phase diagram of $SG_1$. We appeal to a variational method to study 
this following Ref.~\onlinecite{diehl}. 
The variational treatment for the sine-Gordon model $SG_1$
replaces the action
\be
S_{SG} = \int d^2\br \left[ \frac{g}{2} (\nabla \phi)^2 - v_1
\cos\phi \right]
\ee
by the variational action
\be
S_0 = \int d^2\br \left[ \frac{\tilde{g}}{2} 
(\nabla \phi)^2 + \frac{1}{2} m \phi^2 \right]
\ee
with the understanding that when $g$ gets large enough, $v_1$ will
tend to pin the field $\phi$ to integer values leading to
a mass for $\phi$ fluctuations. 

At leading order, the variational free energy
\be
f_{\rm var} \approx f_0 + \la (S_{SG} - S_0) \ra_0.
\ee
is minimized for a mass $m$ which satisfies the self-consistency
condition
\bea
\tilde{g}&=&g\\
m&=&v_1 (\frac{m}{m+4\pi g})^{1/8\pi g}
\eea
It is easy to show that for $v_1 \lesssim 0.5$, the mass is zero for
$g < g_{\rm KT} = 1/8\pi$ while it increases continuously for 
$g > 1/8\pi$. This is the regime in which the continuous KT 
transition obtains in this model. At larger $v_1 \gtrsim 0.5$, 
the mass jumps discontinuously from zero to a nonzero value at
a transition point $g_c < 1/8\pi$. This phase diagram is qualitatively
illustrated in Fig.~\ref{fig:pd_suspect}(b).

\subsection{Scenario for the SF-normal thermal transition}
\begin{figure}
\includegraphics[width=3.2in]{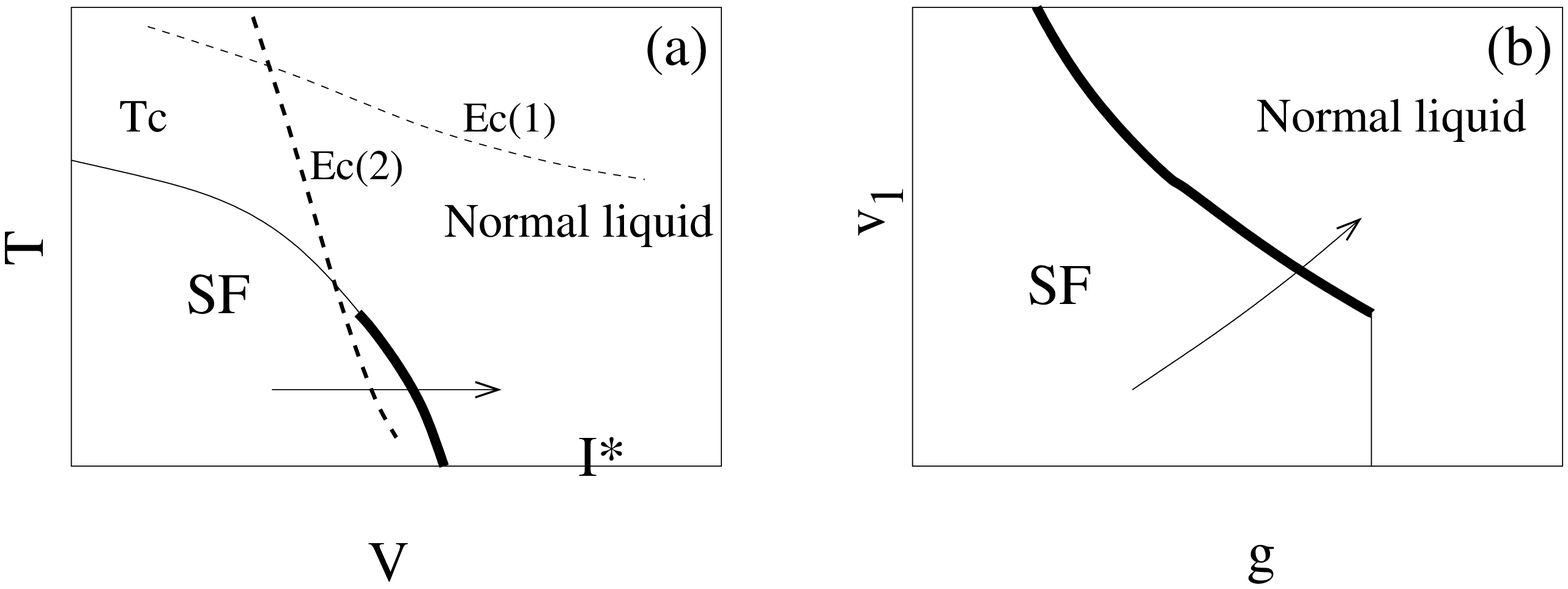}
\caption{\underline{(a)}: Assumed behavior of vortex core energies as a 
function
of $V/t$ in the effective classical model of vortices used to describe
the thermal superfluid-normal phase transition. The solid lines are a
schematic of
the transition curve $T_c(V)$ as obtained from the SSE simulations. 
Light solid line shows
the region where this superfluid-normal transition is of the Kosterlitz
Thouless type. Dark solid line indicates where the transition becomes
first order. The arrow indicates a constant $T$ trajectory going
through the SF-normal transition. \underline{(b)}: The corresponding 
phase diagram in the
usual sine-Gordon model. The trajectory in the left panel translates into
this modified trajectory in the $(g,v_1)$ plane.}
\label{fig:pd_suspect}
\end{figure}

We now appeal to the above results
to understand
the thermal transition from the SF-normal thermal transition.
Far from the quantum phase transition between the
superfluid and the $Z_2$ fractionalized insulator, the thermal
disordering of the superfluid proceeds via the usual KT transition.
As the zero temperature phase approaches the quantum critical point
however, the fugacity of double vortices at finite nonzero $T$
becomes much larger than
the fugacity of single vortices since the SF-${\cal I}^*$ quantum phase 
transition is driven by double-vortex condensation. We model this
situation in our classical Hamiltonian by setting
$E_c(2) \to 0$ near the quantum phase transition, and keeping a 
nonzero and large $E_c(1)$. We expect this effective classical
description to be adequate so long as we are not too close to the
quantum critical point. Fig.~\ref{fig:pd_suspect}(a) illustrates
the expected qualitative behavior of the single and double vortex
core energies, as well as $T_c$, as a function of $V/t$.

Let us imagine a trajectory at fixed temperature but along increasing
$V/t$. In this case, the bare superfluid stiffness decreases with
increasing $V/t$, so that $g$ is increasing monotonically.
Let us assume for simplicity that the bare single vortex core energy
$E_c(1)$ does not change
with $V/t$. However $E_c(2)$ rapidly drops near the quantum phase
transition upon increasing $V/t$. From our earlier discussion
the large $v_2$ then leads to an effectively larger $v_1$. Thus,
constant temperature trajectories in the $(V/t,T/t)$ plane are expected
to translate into trajectories depicted in Fig.~\ref{fig:pd_suspect}(b) 
in the $(g,v_1)$ plane. We argue that this may be responsible for the 
observed first order thermal transition from the superfluid to the normal 
phase close to the quantum phase transition point.
This first order transition must be
accompanied by a nonuniversal jump in the superfluid stiffness which is 
larger than that predicted by KT theory. This is consistent with 
numerical observations presented in the previous sections.

%------------------------------------------------------------------------------
\section{Conclusion}
To summarize, we have studied the zero and finite temperature
phase diagram of a model of hard
core bosons with local interactions which exhibits a topologically
ordered $Z_2$ insulating phase at zero temperature. 
In magnetic language, this is equivalent
to finding a quantum spin liquid phase of a $S=1/2$ quantum magnet.
We have presented a number of numerical results, and some analytical
arguments, in support of this identification. We have also studied
the finite temperature phase diagram and identified a ``cooperative
paramagnet'' regime, and seen that the superfluid to ``cooperative
paramagnet'' transition is a first order transition rather than a 
BKT transition. Further work is needed to see if there is any
connection between spin liquids found in simple model Hamiltonians,
such as the one studied here, and the experimentally observed spin 
liquids in quantum magnets on kagome lattices \cite{spinliq:expt}.

%------------------------------------------------------------------------------
\section*{Acknowledgments}

We acknowledge support from NSERC (SVI, YBK, AP), CRC, CIAR and
KRF-2005-070-C00044 (SVI, YBK), and an Alfred. P. Sloan Foundation
Fellowship (AP). We thank L. Balents, R. G. Melko, and B. Seradjeh for useful 
discussions.


\begin{thebibliography}{99}
\bibitem{footnote.defects}
Such defects correspond to particle-hole pairs 
in real space, which are necessarily bound to each other since the system is 
insulating and charges cannot propagate freely in the ground state.
\end{thebibliography}
\end{document}